# BLOCKCHAIN-BASED SMART CONTRACTS : A SYSTEMATIC MAPPING STUDY

Maher Alharby[1,2] and Aad van Moorsel[1]

[1]School of Computing Science, Newcastle University, Newcastle, UK
[2]College of Computer Science and Engineering, Taibah University,
Medina, KSA

## ABSTRACT

*An appealing feature of blockchain technology is smart contracts. A smart contract is executable code that runs on top of the blockchain to facilitate, execute and enforce an agreement between untrusted parties without the involvement of a trusted third party. In this paper, we conduct a systematic mapping study to collect all research that is relevant to smart contracts from a technical perspective. The aim of doing so is to identify current research topics and open challenges for future studies in smart contract research. We extract 24 papers from different scientific databases. The results show that about two thirds of the papers focus on identifying and tackling smart contract issues. Four key issues are identified, namely, codifying, security, privacy and performance issues. The rest of the papers focuses on smart contract applications or other smart contract related topics. Research gaps that need to be addressed in future studies are provided.*

## 1. INTRODUCTION

Transactions between parties in current systems are usually conducted in a centralised form, which requires the involvement of a trusted third party (e.g., a bank). However, this could result in security issues (e.g., single point of failure) and high transaction fees. Blockchain technology has emerged to tackle these issues by allowing untrusted entities to interact with each other in a distributed manner without the involvement of a trusted third party. Blockchain is a distributed database that records all transactions that have ever occurred in a network. Blockchain was originally introduced for Bitcoin (a peer-to-peer digital payment system), but then evolved to be used for developing a wide range of decentralised applications. An appealing application that can be deployed on top of blockchain is smart contracts.

A smart contract is executable code that runs on the blockchain to facilitate, execute and enforce the terms of an agreement between untrusted parties. It can be thought of as a system that releases digital assets to all or some of the involved parties once the pre-defined rules have been met [1]. Compared to traditional contracts, smart contracts do not rely on a trusted third party to operate, resulting in low transaction costs. There are different blockchain platforms that can be utilised to develop smart contracts, but Ethereum is the most common one. This is because Ethereum's language supports Turing-completeness feature that allows creating more advanced and customised contracts. Smart contracts can be applied to different applications (e.g., smart properties, e-commerce and music rights management).

The main aim of this study is to identify the research topics that have been carried out about blockchain-based smart contracts and current challenges that need to be addressed in future studies. To achieve this aim, we selected a systematic mapping study as the methodology for our study. We followed the systematic mapping process presented in [2] to search for relevant papers in scientific databases and to produce a map of current smart contract research. The produced map could help researchers identify gaps for future studies. The focus of our study is to only explore smart contract studies from a technical point of view.

The structure of this paper is as follows. Section 2 discusses background information about blockchain and smart contracts technologies. It also discusses several smart contract platforms and potential applications. Section 3 describes the research methodology adopted for our study. Section 4 presents the results of searching and screening for relevant papers and the results of classifying smart contract topics. Section 5 discusses the results and answers the research questions of the study. Section 6 concludes the paper.

## 2. BACKGROUND

This section presents general background information about blockchain and smart contracts technologies. It also discusses some blockchain platforms that support the development of smart contracts. Finally, it provides some potential use cases for smart contracts.

### 2.1. Blockchain Technology

A blockchain is a distributed database that records all transactions that have ever occurred in the blockchain network. This database is replicated and shared among the network's participants. The main feature of blockchain is that it allows untrusted participants to communicate and send transactions between each other in a secure way without the need of a trusted third party. Blockchain is an ordered list of blocks, where each block is identified by its cryptographic hash. Each block references the block that came before it, resulting in a chain of blocks. Each block consists of a set of transactions. Once a block is created and appended to the blockchain, the transactions in that block cannot be changed or reverted. This is to ensure the integrity of the transactions and to prevent double-spending problem.

Cryptocurrencies have emerged as the first generation of blockchain technology. Cryptocurrencies are basically digital currencies that are based on cryptographic techniques and peer-to-peer network. The first and most popular example of cryptocurrencies is Bitcoin. Bitcoin [3] is an electronic payment system that allows two untrusted parties to transact digital money with each other in a secure manner without going through a middleman (e.g., a bank). Transactions that occurred in the network are verified by special nodes (called miners). Verifying a transaction means checking the sender and the content of the transaction. Miners generate a new block of transactions after solving a mathematical puzzle (called Proof of Work) and then propagate that block to the network. Other nodes in the network can validate the correctness of the generated block and only build upon it if it was generated correctly. However, Bitcoin has limited programming capabilities to support complex transactions. Bitcoin, thus, does not support the creation of complex distributed applications on top of it.

Other blockchains such as Ethereum have emerged as the second generation of blockchain to allow building complex distributed applications beyond the cryptocurrencies. Smart contracts, which will be discussed in the following section, are considered as the main element of this generation [4]. Ethereum blockchain is the most popular blockchain for developing smart contracts. Ethereum is a public blockchain with a built-in Turing-complete language to allow writing any smart contract and decentralised application.

There are two types of blockchain, namely, public and private blockchain [5]. In a public blockchain, any anonymous user can join the network, read the content of the blockchain, send a new transaction or verify the correctness of the blocks. Examples of public blockchains are Bitcoin, NXT and Ethereum. In a private blockchain, only users with permissions can join the network, write or send transactions to the blockchain. A company or a group of companies are usually responsible for giving users such permissions prior to joining the network. Examples of private blockchains are Everledger, Ripple and Eris.

## 2.2. Smart Contracts

A smart contract is executable code that runs on the blockchain to facilitate, execute and enforce the terms of an agreement. The main aim of a smart contract is to automatically execute the terms of an agreement once the specified conditions are met. Thus, smart contracts promise low transaction fees compared to traditional systems that require a trusted third party to enforce and execute the terms of an agreement. The idea of smart contracts came from Szabo in 1994 [6]. However, the idea did not see the light till the emergence of blockchain technology. A smart contract can be thought of as a system that releases digital assets to all or some of the involved parties once arbitrary pre-defined rules have been met [1]. For instance, Alice sends X currency units to Bob, if she receives Y currency units from Carl.

Many different definitions of a smart contract have been discussed in the literature. In [7], the author classified all definitions into two categories, namely, smart contract code and smart legal contract. Smart contract code means "code that is stored, verified and executed on a blockchain" [7]. The capability of this smart contract depends entirely on the programming language used to express the contract and the features of the blockchain. Smart legal contract means code to complete or substitute legal contracts. The capability of this smart contract does not depend on the technology, but instead on legal, political and business institutions. The focus of this study will be on the first definition, which is smart contract code.

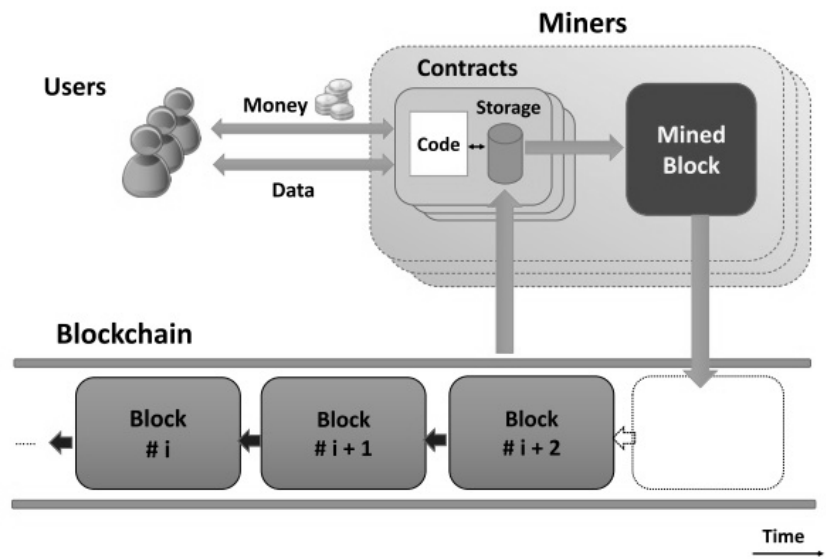

Figure 1.  Smart contract system [8].

A smart contract has an account balance, a private storage and executable code. The contract's state comprises the storage and the balance of the contract. The state is stored on the blockchain and it is updated each time the contract is invoked. Figure 1 depicts the smart contract system.

Each contract will be assigned to a unique address of 20 bytes. Once the contract is deployed into the blockchain, the contract code cannot be changed. To run a contract, users can simply send a transaction to the contract's address. This transaction will then be executed by every consensus node (called miners) in the network to reach a consensus on its output. The contract's state will then be updated accordingly. The contract can, based on the transaction it receives, read/write to its private storage, store money into its account balance, send/receive messages or money from users/other contracts or even create new contracts.

There are two types of smart contracts, namely, deterministic and non-deterministic smart contracts [9]. A deterministic smart contract is a smart contract that when it is run, it does not require any information from an external party (from outside the blockchain). A non-deterministic smart contract is a contract that depends on information (called oracles or data feeds) from an external party. For example, a contract that requires the current weather information to be run, which is not available on the blockchain.

### 2.3. Platforms for Smart Contracts

Smart contracts can be developed and deployed in different blockchain platforms (e.g., Ethereum, Bitcoin and NXT). Different platforms offer distinctive features for developing smart contracts. Some platforms support high-level programming languages to develop smart contracts. We will only focus on three public platforms in this section.

- Bitcoin [3] is a public blockchain platform that can be used to process cryptocurrency transactions, but with a very limited compute capability. Bitcoin uses a stack-based bytecode scripting language. The ability of creating a smart contract with rich logic using Bitcoin scripting language is very limited [10]. In Bitcoin, a simple logic that requires multiple signatures to sign a single transaction before confirming the payment is possible. However, writing contracts with complex logic is not possible due to the limitations of Bitcoin scripting language. Bitcoin scripting language, for example, neither supports loops nor withdrawal limits [1]. To implement a loop, the only possible way is by repeating the code many times, which is inefficient.

- NXT is a public blockchain platform that includes built-in smart contracts as templates [10]. NXT only allows developing smart contracts using those templates. It does not, however, allow customized smart contracts due to the lack of Turing-completeness in its scripting language.

- Ethereum [1,11] is a public blockchain platform that can support advanced and customized smart contracts with the help of Turing-complete programming language. Ethereum platform can support withdrawal limits, loops, financial contracts and gambling markets. The code of Ethereum smart contracts is written in a stack-based bytecode language and executed in Ethereum Virtual Machine (EVM). Several high-level languages (e.g., Solidity, Serpent and LLL) can be used to write Ethereum smart contracts. The code of those languages can then be compiled into EVM bytecodes to be run. Ethereum currently is the most common platform for developing smart contracts.

### 2.4. Smart Contract Applications

There are various possible applications where smart contracts can be applied to. Some of these applications are as follows:

- Internet of Thing and smart property [12]: there are billions of nodes that are sharing data between each other through the Internet. A potential use case of blockchain-based smart contracts is to allow those nodes to share or access different digital properties without a trusted third party. There are various companies that investigate this use case. For example, Slock.it is a German company that utilises Ethereum-based smart contracts for renting, selling or sharing anything (e.g, selling a car) without the involvement of a trusted third party.

- Music rights management [13]: a potential use case is to record the ownership rights of a music in the blockchain. A smart contract can enforce the payment for music owners once a music is used for commercial purposes. It also ensures the payment is being distributed between the music's owners. Ujo is a company that investigates the use of blockchain-based smart contracts in the music industry.

- E-commerce: a potential use case is to facilitate the trade between untrusted parties (e.g., seller and buyer) without a trusted third party. This would result in reduction of trading costs. Smart contracts can only release the payment to the seller once the buyer is satisfied with the product or service they received [14].

There are other possible applications such as e-voting, mortgage payment, digital right management, motor insurance, distributed file storage, identity management and supply chain.

## 3. RESEARCH METHODOLOGY

We selected the systematic mapping study presented in [2] as the research methodology for our study to explore studies related to smart contracts. The results of this systematic mapping study would allow us to identify and map research areas related to smart contracts. In addition, it would allow us to identify research gaps that need to be considered for future studies. The process for the systematic mapping study falls into five steps as depicted in Figure 2.

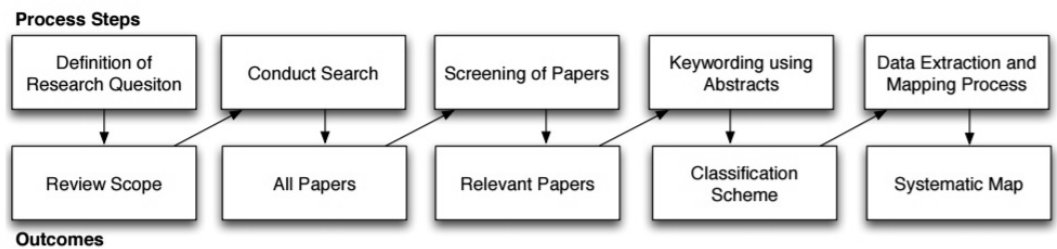

Figure 2.  Steps of the systematic mapping study [2].

**Definition of research questions:**

This step is to identify the research questions the study is aiming to answer. For our study, we defined the following research questions:

RQ1. What are the current research topics on smart contracts?

RQ2. What are the current smart contract applications?

RQ3. What are the research gaps that need to be addressed in future studies?

**Conducting the search:**

This step is to search and to find all scientific papers that are related to the research topic, which is smart contracts. For our study, we decided to select the term 'smart contract' as the main keyword to search for papers. We selected this term because we wanted to narrow down the focus of our study to only cover smart contract related works. After identifying the keyword for the searching process, we selected the scientific databases to conduct our search. We selected IEEE Explore, ACM Digital Library, ScienceDirect, Springer, Ebsco and Scopus. Our focus was to only include high quality papers published in conferences, journals, workshops, symposiums and books.

**Screening for relevant papers:**

This step is to search for papers that are relevant to our research questions. We followed the same approach as in [15] to look for relevant papers. We first tried to exclude papers that were irrelevant to our study based on their titles. If we were unable to decide on a paper, we would go a step further by examining its abstract. We also used exclusion criteria to screen each paper. We excluded: (1) non-English papers, (2) papers without full text available, (3) papers that utilised smart contracts in fields other than computer science, (4) redundant papers and (5) articles, newsletters and grey literature.

**Key-wording using abstracts:**

This step is to classify all relevant papers using the key-wording technique described in [15]. We first read the abstract of each paper to identify the most important keywords and the main contribution. Those keywords were then used to classify papers into various categories. After classifying all papers, we read the papers and made changes to the classification when necessary.

**Data extraction and mapping process:**

This process is to gather all the required information to address the research questions of this study. We gathered different data items from each paper. These data items embrace the main aims and contributions of papers.

## 4. STUDY RESULTS

This section discusses the results of the systematic mapping study that we conducted on smart contracts. We first discuss the results of searching and screening for relevant papers. Then, we discuss the results of the classification process.

### 4.1 Searching and Screening Results

Searching and screening for relevant papers are two steps of the systematic mapping study that we discussed in Section 3. The results of these steps are as follows. In the searching phase, we looked for all papers using the term 'smart contract' in different scientific databases. We gathered 154 papers in total (as on 5 May 2017). In the screening phase, we first excluded irrelevant papers based on their titles and/or their abstracts (we excluded 109 irrelevant papers). There are two reasons why we had a high number of excluded papers. First, many papers were irrelevant to our study, since our focus was to explore smart contracts from a technical perspective. For instance, many papers discussed the topic from an economic or legal point of view. Another reason is that some excluded papers were about cryptocurrencies or blockchain in general, which do not contribute to our research questions. After that, 17 papers were removed as they were duplicates, resulting in 28 papers. Among the 28 papers, four papers were excluded as they only discuss general information about smart contract and how it works, without providing any useful contribution. Thus, we only selected 24 papers to conduct our systematic mapping study. Figure 3 summaries the results of searching and screening for relevant papers.

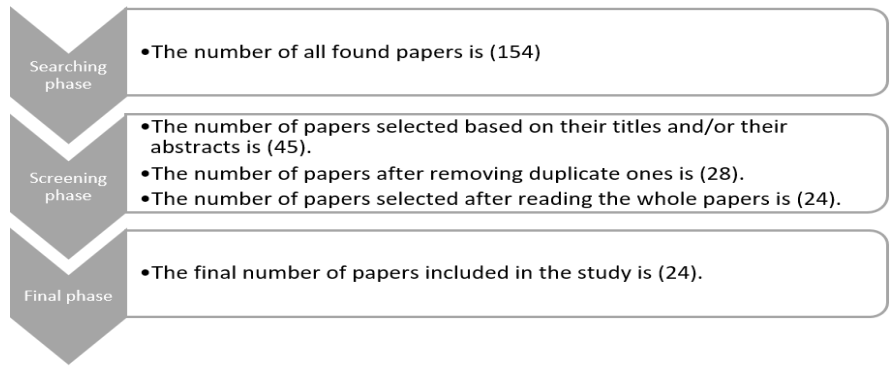

Figure 3.  Searching and screening results.

## 4.2 Classification Results

By applying the Key-wording technique that we discussed in Section 3, we classified the papers into two categories, namely, smart contract issues and other smart contract related topics. We found about two thirds of the papers fall into smart contract issues category. We classified those issues into four categories, namely, codifying, security, privacy and performance issues. *Codifying issues* mean challenges that are related to the development of smart contracts. *Security issues* mean bugs or vulnerabilities that an adversary might utilise to launch an attack. *Privacy issues* mean issues related to disclosing contracts information to the public. *Performance issues* mean issues that affect the ability of blockchain systems to scale. Table 1 summaries the identified issues and the proposed solutions. For other smart contract related topics category, there are nine papers that developed smart contract applications or reported about other topics (e.g., the combination of smart contract and The Internet of Thing).

Table 1.  Smart contract issues and the proposed solutions.

| Smart contract issues | | Proposed solutions |
|---|---|---|
| **Codifying issues** | Difficulty of writing correct smart contracts [8,16,17,18]. | • Semi-automation of smart contracts creation [18].<br>• Use of formal verification methods [16,17].<br>• Education (e.g., online tutorials) [8]. |
| | Inability to modify or terminate smart contracts [19]. | • A set of standards for modifying/terminating smart contracts [19]. |
| | Lack of support to identify under-optimised smart contracts [20]. | • Use of 'GASPER' tool [20]. |
| | Complexity of programming languages [21]. | • Use of logic-based languages [21]. |
| **Security issues** | Transaction-ordering dependency vulnerability [22,23]. | • Use of 'SendIfReceived' function [22].<br>• Use of a guard condition [23].<br>• Use of 'OYENTE' tool [23]. |
| | Timestamp dependency vulnerability [23]. | • Use block number as a random seed instead of using timestamp [23].<br>• Use of 'OYENTE' tool [23]. |
| | Mishandled exception vulnerability [23]. | • Check the returned value [23].<br>• Use of 'OYENTE' tool [23]. |
| | Re-entrancy vulnerability [23]. | • Use of 'OYENTE' tool [23]. |

| | Criminal smart contract activities [24]. | • NA. |
|---|---|---|
| | Lack of trustworthy data feeds 'Oracles' [25]. | • Use of 'Town Crier (TC)' tool [25]. |
| **Privacy issues** | Lack of transactional privacy [26]. | • Use of 'Hawk' tool [26].<br>• Use of encryption techniques [27]. |
| | Lack of data feeds privacy [25]. | • Use of 'Town Crier (TC)' tool [25].<br>• Use of encryption techniques [25]. |
| **Performance issues** | Sequential execution of smart contracts [28]. | • Parallel execution of smart contracts [28]. |

## Codifying issues

From the literature, we found four issues that might face developers during writing smart contracts, namely, the difficulty of writing correct contracts, the inability to modify or terminate contracts, the lack of support to identify under-optimised contracts and the complexity of programming languages.

The first one is the difficulty of writing correct smart contracts [8,16,1718]. Correctness of smart contracts in this context means contracts that are functioning as intended by their developers. The reason why it is important to have correct smart contracts is because those contracts have valuable currency units [8,16]. Thus, if a smart contract was not executed as intended, some of its currency units would disappear. An example that illustrates this is the Distributed Autonomous Organisation (DAO) attack, which led to over 60 million US dollars being moved into an adversary account [23].

In an attempt to tackle this issue, three solutions were identified from the literature. The first solution is to semi-automate the creation of smart contracts [18] to ease the process of writing smart contracts. Semi-automation means the translation of human-readable contract representations to smart contract rules. The second solution is to provide developers with guidelines to aid them write correct contracts. Delmolino et al. [8], released online materials (e.g., a tutorial) to help developers write correct smart contracts. The last solution is the adoption of formal verification techniques to detect unintended behaviours of smart contracts [16,17]. This can help developers recognise those behaviours before posting their contracts to the blockchain. Bhargavan et al. [16] utilised formal methods to analyse and verify the correctness of smart contracts, while Bigi et al. [17] went a step further by combining formal methods with game theory techniques to validate smart contracts.

The second issue is the inability to modify or terminate smart contracts [19]. Due to the immutability feature of blockchain, smart contracts cannot be changed or terminated after deploying it into the blockchain. This is different from legal law which allows the rules to be modified or terminated. In an attempt to tackle this issue, Marino et al. [19] presented a set of standards to allow smart contracts to be changed or terminated. Such standards are taken from legal contracts and then defined to fit in the context of smart contracts. Those standards were then applied to Ethereum-based smart contracts to prove their success. For details about those standards and how can be applied to Ethereum-based smart contracts, we refer the reader to [19].

The third one is the lack of support to identify under-optimised smart contracts [20]. To run a smart contract, each computational or storage operation in the contract costs some money. An under-optimised smart contract is a contract that contains unnecessary or expensive operations. Such operations result in a high cost at the user's side. In an attempt to tackle this issue, Chen et al. [20] identified seven programming patterns (e.g., unnecessary and expensive operations in a loop) in smart contracts which lead to unnecessary extra costs. They also proposed ways to enhance the optimisation of those patterns to reduce the overall cost of executing smart contracts. They proposed and developed a tool called 'GASPER' to detect contracts that suffer from those patterns. They used the tool to examine current Ethereum smart contracts and found most of them suffer from such patterns.

The last issue is the complexity of smart contract programming languages [21]. Current smart contracts are based on procedural languages such as Solidity. In a procedural language, the code is executed as a sequence of steps. Thus, programmers must specify what should be done and how to do it. This makes the task of writing smart contracts in those languages cumbersome and error prone [21]. In an attempt to tackle this issue, Idelberger et al. [21] proposed to utilise logic-based languages instead of procedural languages. In logic-based languages, programmers do not necessarily have to specify the sequence of steps for a contract. This will ease the complexity of writing smart contracts. However, algorithms for logic-based languages are expensive and inefficient.

## Security issues

From the literature, we found six security issues, namely, transaction-ordering dependency, timestamp dependency, mishandled exception, criminal activities, re-entrancy and untrustworthy data feeds. In addition to these issues, Atzei et al. [29] surveyed several vulnerabilities in Ethereum smart contracts.

The first issue is transaction-ordering dependency [22,23]. This problem occurs when two dependent transactions that invoke the same contract are included in one block. The order of executing transactions relies on the miner. However, an adversary can successfully launch an attack if those transitions were not executed in the right order. For example, assume there is a puzzle contract that incentives the user who solves the puzzle. A malicious owner is listening to the solutions provided by the users. Once a user submitted a correct solution to the puzzle (Tu), the malicious owner sends a transaction (To) to update the contract's reward (e.g., reduce the reward) right away. Those two transactions (To and Tu) might be included in the same block by chance. If the miner executed To before Tu, the user would get a lower reward and the malicious owner would succeed in his attack [23]. To tackle this issue, Natoli et al.[22] suggested the use of Ethereum-based functions (e.g., SendIfReceived) to enforce the order of transactions. Similarly, Luu et al.[23] suggested using a guard condition such that "a contract code either returns the expected output or fails". A tool called 'OYENTE' developed by [23] can be used to detect contracts that are vulnerable to transaction-ordering dependency.

The second issue is timestamp dependency [23]. This problem occurs when a contract uses the block timestamp as a condition to trigger and execute transactions (e.g., sending money). For instance, a game-based contract that takes the block timestamp as a random seed to select the winner. The block timestamp is usually set as the current local time by the miner who generated the block. However, an issue with the timestamp is that a dishonest miner could vary its value by about 15 minutes from the current time, while the block is still accepted by the blockchain system. As the timestamp of a block is not guaranteed to be accurate, contracts that rely on timestamp value are vulnerable to threats by dishonest miners. To tackle this issue, Luu et al.[23] suggested using the block number as a random seed for contracts instead of using the block timestamp. This is because the value of the block number is fixed (miners cannot vary the block number value). To detect contracts that are vulnerable to timestamp dependency, 'OYENTE' tool presented in [23] can be used.

The third issue is mishandled exception vulnerability [23]. This problem occurs when a contract (caller) calls another contract (callee) without checking the value returned by the callee. When calling another contract, an exception (e.g., run out of gas) sometimes raised in the callee contract. This exception, however, might/might not be reported to the caller depending on the construction of the call function. Having not reported an exception might lead to threats as in the KingOfTheEther (KoET) contract [23]. In KoET, an adversary might send a transaction that results in an exception in order to buy the throne from the current king for free. To tackle this issue, Luu et al.[ 23] highlighted the importance of checking the value returned by the callee. In the KoET example, the code can be improved to not release the throne till the payment from the adversary is completed successfully without any exception. The 'OYENTE' tool proposed by [23] can be used to detect mishandled exception vulnerability in smart contracts.

The fourth issue is re-entrancy vulnerability [23]. This problem occurs when an attacker utilises a recursive call function to conduct multiple repetitive withdrawals, while their balances are only deduced once. In June 2016, an attacker utilised the re-entrancy vulnerability in the Decentralised Autonomous Organisation (DAO) to steal over 60 million US dollars [23]. Luu et al. [23] developed a tool called 'OYENTE' to detect this vulnerability.

The fifth issue is criminal activities. Jules et al. [24] highlighted the feasibility of constructing three different types of criminal activities in smart contract systems, namely, "leakage/sale of secret documents, theft of private keys and calling-card crimes, a broad class of physical-world crimes (murder, arson, etc.)" [24]. These crimes can be implemented efficiently in the Ethereum blockchain by utilising cryptographic techniques as follows. Leakage of secret documents can be achieved with the support of Serphent (an Ethereum scripting language). Theft of private keys can be achieved using Succinct Non-interactive ARgument of Knowledge (SNARKs) cryptographic primitives. Authenticated data feeds, which is data from an external party, can facilitate the calling-card crimes. The authors of [24], however, did not attempt to tackle those crime activities, but instead, they highlighted the importance of constructing safeguards against such activities.

The last issue is the lack of trustworthy data feeds (oracles) [25]. As we explained in Section 2.2, some smart contracts require information (data feeds) from outside the blockchain. The problem is that there is no guarantee that the information provided by an external source is trustworthy. In an attempt to tackle this issue, Zhang et al. [25] built a Town Crier (TC) solution that acts as a trusted third party between external sources and smart contracts to provide authenticated data feeds for smart contracts. Figure 5 explains the architecture of TC solution. The TC solution consists of a TC contract that resides on the blockchain and a TC server that resides outside the blockchain. To send a data feeds request, a user contract can send a request to the TC contract, which will then be forwarded to the TC server. The server then communicates with external data sources via HTTPS to get the data feeds. Upon getting the required data feeds, the server will forward those feeds to the TC contract, which will then be forwarded to the user contract.

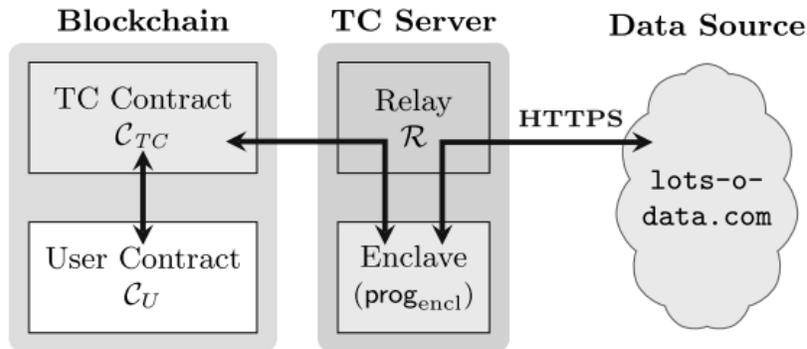

Figure 4. Architecture of TC solution [25].

**Privacy issues**

From the literature, we found two privacy issues, namely, the lack of transactional privacy and the lack of data feeds privacy.

The first issue is the lack of transactional privacy [26,27]. In blockchain systems, all transactions and users' balances are publicly available to be viewed. This lack of privacy could limit the adoption of smart contracts as many people consider financial transactions (e.g., stock trading) as confidential information [26]. To tackle this issue, Kosba et al.[26] built a tool called 'Hawk' that allows developers to write privacy-preserving smart contracts without the need of implementing any cryptography. The tool is responsible for compiling smart contract code to privacy-preserving one. Watanabe et al.[27] proposed to encrypt smart contracts before deploying them to the blockchain. Only participants, who are involved in a contract, can access the contract's content by using their decryption keys.

The second issue is the lack of data feeds privacy [25]. When a contract requires data feeds to operate, it sends a request to the party that provides those feeds. However, this request is exposed to the public as anyone in the blockchain can see it. To tackle this issue, Zhang et al. [25] extend their Town Crier (TC) tool to support private requests. A contract can encrypt the request using the TC's public key, before sending the request. Upon receiving the encrypted request, the TC can decrypt it using its private key. Thus, this would guarantee that the content of the request is kept secret from other users/contracts in the blockchain.

**Performance issues**

From the literature, we only found one performance issue, which is the sequential execution of smart contracts [28]. In blockchain systems, smart contracts are executed sequentially (e.g., one contract at a time). However, this would affect the performance of the blockchain systems negatively as the number of smart contracts that can be executed per second will be limited. With the growing number of smart contracts in the future, the blockchain systems will not be able to scale. Vukolić [28] suggested to execute smart contracts in parallel as long as they are independent (e.g., "do not update the same variables" [28]). By doing so, the performance of blockchain systems would be improved as more contracts can be executed per second.

**Other topics**

Apart from smart contract issues, we found nine papers from the literature that propose smart contract applications or discuss other smart contract related topics.

There are four smart contract applications proposed in the literature, namely, trading and fair exchange, identity management, Internet of Thing and agreements establishment applications. For trading and fair exchange, Bogner et al. [30] developed a smart contract application on top of the Ethereum blockchain to allow untrusted participants to share everyday objects (e.g., rent devices). For identity management, Al-Bassam et al. [31] built a system called `SCPKI' on top of the Ethereum blockchain to overcome the limitations (e.g, centralisation and lack of transparency) of the Public Key Infrastructure. This system allows entities to manage their identities in a transparent way without the involvement of a trusted third party such as central authorities. For the Internet of Thing, Huh et al. [32] used Ethereum smart contracts to define and manage the behaviours of a few devices under specified conditions. For example, an air conditioner that switches to energy saving mode when the usage of electricity reaches 170 KW. For agreements establishment, Carrillo et al. [33] developed an application that allows two untrusted parties (e.g., consumer and provider) to negotiate and then establish an agreement as a contract.

In addition to smart contract applications, there are different topics that were discussed in the literature. In [12], the authors discussed how the combination of blockchain-based smart contracts with the Internet of Thing could be powerful in terms of facilitating the sharing of services. In [9], the authors discussed the possibility of applying blockchain-based smart contracts for licensing management. For example, the use of smart contracts to control the license of software products. In [14], the authors investigated the possibility of creating complex smart contracts without relying on scripts. In [34], the authors proposed a new consensus method called 'credibility' for contracts management (e.g., digital right management) to avoid the limitations of existing consensus methods. In [35], the authors proposed a semantic index approach to search for information in the Ethereum blockchain.

## 5. DISCUSSION

This section discusses the study results and answers the research questions that we defined in Section 3.

**RQ1: What are the current research topics on smart contracts?**

The results of this systematic mapping study showed that most of the current research on smart contracts is about identifying and tackling smart contract issues. Four different issues were identified, namely, codifying, security, privacy and performance issues. Codifying and security issues were among the most discussed issues. This is because smart contracts store valuable currency units and any security breach or coding error could result in losing money. The identified codifying issues are the difficulty of writing correct codes, the inability to modify or terminate contracts, the lack of support to identify under-optimised contracts and the complexity of programming languages. The identified security issues are transaction-ordering dependency, timestamp dependency, mishandled exception, re-entrancy, untrustworthy data feeds and criminal activities. The identified privacy issues are the lack of transactional privacy and the lack of data feeds privacy. The identified performance issue is the sequential execution of smart contracts. Although there are some proposed solutions to tackle these issues, some of them are only abstract ideas without including any concrete evaluation. A few others are still not tackled yet. For example, the solution proposed by [21] is only a suggestion to use alternative programming languages without any implementation. Criminal activities identified by [24] are still not overcome yet.

Other research proposed smart contract applications or studied other smart contract related topics. The proposed applications are trading and fair exchange, identity management, Internet of Thing and agreements establishment. The studied topics are combining smart contracts with the Internet of Thing and licensing management, studying scripting languages for smart contracts, proposing new consensus methods and proposing an indexing approach to search for useful information in blockchain systems.

**RQ2: What are the current smart contract applications?**

Smart contract applications are solutions that have been developed on top of blockchain technology. We identified some smart contract applications developed on top of the Ethereum blockchain. Those applications are to allow untrusted participants to share everyday objects, establish an agreement as a contract, manage their identities and control the behaviours of the Internet of Thing devices. Furthermore, we identified other applications that were built as a smart contract tool on top of the blockchain to detect or tackle codifying, security and privacy issues. Some of these tools are 'GASPER', 'OYENTE', 'HAWK' and 'Town Crier'.

**RQ3: What are the research gaps that need to be addressed in future studies?**

From this systematic mapping study, we identified a number of research gaps in smart contract research that can be studied by future research. The methodologies used to identify those gaps are as follows. First, observing issues or limitations from the papers included in this study (e.g., gaps number 2, 3 and 5). Second, recognising issues that were highlighted by the papers included in this study, but still are not solved yet (e.g., gaps number 1 and 4).

The first one is the lack of studies on scalability and performance issues. The sequential execution of smart contracts affects the ability of blockchain systems to scale as we discussed in Section 4.2. With the growing number of smart contracts in the future, this issue will increase further. The author of [28] described a very high-level solution, which is parallel execution of contracts, without any concrete evaluation. Parallel execution of contracts faces a challenge in how to execute contracts that depend on each other at the same time. It is, therefore, essential to conduct research on identifying and tackling performance issues to ensure the ability of blockchain to scale.

The second gap is that almost all current research is discussing smart contracts on the Ethereum blockchain, although there are some other blockchains (e.g., NXT and Eris) that can support the creation of smart contracts. Different blockchains have distinctive features and advantages. Thus, future research might investigate different implementations of blockchain to deploy and run smart contracts.

The third gap is the small number of smart contract applications. Although the concept of smart contract has gained a lot of attention, there are only a few applications developed by the literature. This is because smart contract concept is still in its infancy stage. Banasik et al.[14] claimed that smart contracts are not widely common in practice. For future research, therefore, researchers could consider studying various potential applications such as e-commerce and cloud storage.

The fourth gap is the lack of research on tackling criminal activities in smart contracts. The author of [24] only identified three types of criminal activities that can be conducted on smart contracts without proposing any solution to them. Thus, future research could focus on identifying more types of criminal activities and proposing solutions to overcome them.

The last gap is the lack of high quality peer-reviewed research on smart contracts. Most of the research is conducted as blog articles or grey literature without providing great contributions. There is, therefore, a need for high quality publications on smart contracts.

## 6. CONCLUSION

Blockchain technology is a distributed database that records all transactions that have ever occurred in the network. The main feature of blockchain is that it allows untrusted parties to communicate between each other without the need of a trusted third party. Different distributed applications beyond cryptocurrencies can be deployed on top of blockchain. One of these applications is smart contracts, which are executable codes that facilitate, execute and enforce an agreement between untrusted parties. Ethereum is currently the most common blockchain platform for developing smart contracts, although there are some other available platforms.

To understand current topics on smart contracts, we decided to conduct a systematic mapping study. The main aim of this systematic mapping study was to identify and map research areas related to smart contracts. By doing so, we were able identify research gaps that need to be addressed in future studies. The focus of this study was on smart contracts from a technical point of view. Thus, we excluded studies with different perspectives (e.g., papers with an economic perspective). We extracted 24 papers from different databases. We found that most papers identifying and tackling issues on smart contracts. We grouped these issues into four categories, namely, codifying, security, privacy and performance issues. The rest of the papers focuses on proposing smart contract applications or discussing other smart contract related topics.

In this paper, we presented a few research gaps in smart contract research that need to be addressed in future studies. The identified gaps are the lack of studies on scalability and performance issues, the lack of studies on deploying smart contracts on different blockchain platforms other than Ethereum, the small number of the proposed smart contract applications, the lack of studies on criminal activities in smart contracts and the lack of high quality research on smart contracts. These identified gaps could be studied by researchers as future works.